\documentclass [12pt] {article}

\newcommand {\hyperEnableP} {}

\newcommand {\theTitle} {
	Conformal invariance on orbifolds and excitations of singularity
}
\newcommand {\thePreprintTitle} {
	CONFORMAL INVARIANCE ON ORBIFOLDS\\ AND\\ EXCITATIONS OF SINGULARITY
}

\newcommand {\theAbstract} {
We study conformal field theory on two-dimensional orbifolds and show
this to be an effective way to analyze physical effects of geometric
singularities with angular deficits.  They are closely related to
boundaries and cross caps. Representatives classes of singularities can
be described exactly using generalizations of boundary states.  From
this we compute correlation functions and derive the spectra of
excitations localized at the singularities.
}

\newcommand {\thePreprintNumbers} {hep-th/0701056 \cr SIAS-CMTP-06-5}

\newcommand {\theAuthorList}{Zheng YIN}

\newcommand {\theAuthorFootnote}{yinzheng at ustc point edu dot cn}

\newcommand {\theAddress} {
Center for Mathematics and Theoretical Physics\\
Shanghai Institute for Advanced Studies\\
99 Xiupu Rd, Shanghai, China, 201315
}

\newcommand {\theShortAddress} {
	USTC-Shanghai Institute for Advanced Studies\\
	99 Xiupu Road; Shanghai, PC 201315; China
}

\newcommand {\theDate} {December 2006}

\usepackage [final] 
	{graphicx}

\usepackage{amssymb}
\newcommand {\ignore} [1] {}
\newcommand {\ignoreTwo} [2] {}

\newcommand {\ZZ} {\mathbb Z}

\newcommand {\tL} {{\tilde L}}

\newcommand {\tDelta} {{\tilde \Delta}}
\newcommand {\tlambda} {{\tilde \lambda}}

\newcommand {\ket} [1] {\left|#1\!\right>}

\newcommand {\BSket}[1] [{ }] {\left|\left.#1\right>\!\right)}

\newcommand {\beq} {\begin {equation}}
\newcommand {\eeq} {\end {equation}}

\newcommand {\beqn} {\begin {displaymath}}
\newcommand {\eeqn} {\end {displaymath}}

\newcommand {\beqar} {\begin {eqnarray}}
\newcommand {\eeqar} {\end {eqnarray}}

\newcommand {\beqarn} {\begin {eqnarray*}}
\newcommand {\eeqarn} {\end {eqnarray*}}

\newcommand {\nonum} {\nonumber}
\newcommand {\nono} {\nonumber \\ {}}

\newcommand{\abs}[1]{\left|{#1}\right|}	% | . |

\newcommand {\csp} {\;\;}

\newcommand {\eqr} [1]  {{(eq. \ref {eq:#1})}}
\newcommand{\brak}[1]{\left[{#1}\right]}	% [ . ]
\newcommand{\bracP}[1]{\left({#1}\right)}	% ( . )
\newcommand{\bracA}[1]{\left\langle{#1}\right\rangle}	% < . >

\newcommand{\comm}[1]{{[#1]}}	% [ . ]

\newcommand {\lieg} {{\mathfrak {g}}}

\newcommand {\figRef} [1] {{fig. \ref {fig:#1}}}

\newcommand {\FigRef} [1] {{Fig. \ref {fig:#1}}}

\newcommand {\tz} {{\tilde z}}

\newcommand {\td} {{\tilde d}}

\newcommand {\tbl} {{\tilde \bl}}

\newcommand {\tlh} {{\tilde h}}

\newcommand {\bl} {l}

\newcommand {\compactTitle} {
	\title{\theTitle}
	\author{\theAuthorList}
	\preprint {\thePreprintNumbers}
	\affiliation{\theShortAddress}
	\date{\theDate}
	\begin{abstract}
	\theAbstract
	\end{abstract}

	\maketitle
}

\usepackage {ifthen}

\ifthenelse {\isundefined {\draftFormatP}}
	{\ifthenelse {\isundefined {\hyperEnableP}}
		{\def\href#1#2{#2}}
		{\usepackage [draft=false] {hyperref}}
	 \def\url#1{\texttt{#1}}
	}
	{
	 \ifthenelse {\isundefined {\outputNonpreprintP}}
		{\usepackage [draft=false, backref] {hyperref}}
		{\usepackage [draft=false, backref, pagebackref] {hyperref}}
	}

\ifthenelse {\isundefined {\outputNonpreprintP}} 
{

	\def\sct{\section}

	\usepackage [square,numbers] {natbib}

	\newcommand {\zyEpsf} [3]
		{\begin {figure} [htbp]
			\begin {center}
				\includegraphics {#1}
				\caption {\label {fig:#2}	#3}
			\end {center}
		\end {figure}}

	\parskip 4pt
	\marginparwidth 0pt 
	\marginparsep 0pt 
	\setlength{\textwidth}{164mm} 
	\setlength{\textheight}{210mm}
	\addtolength{\oddsidemargin}{-14mm} 
	\addtolength{\topmargin}{-13mm}
	\addtolength{\footskip}{5mm}
	\setlength{\parskip}{2mm}

	\makeatletter
}
{% for compact only

	\newcommand {\sct} [1] {{\it #1} --}

	\setlength{\topmargin}{0in}

	\newcommand {\zyEpsf} [3] {
		\begin {figure}
			\includegraphics {#1}
			\caption {\label {fig:#2}	#3}
		\end {figure}}

}

\def\cp{Cardy:1988tk}
\def\cardy{Cardy:1989ir}

\def\krr{Kleppe:1988je}

\def\im{Ilderton:2004xq}
\def\iim{Imamura:2005zm}
\def\bh{Brunner:2002em}

\def\as{Angelantonj:2002ct}
\def\bppz{Behrend:1999bn}

\def\polchinskiBook{Polchinski:1998bookBrf}
\def\polchinskiRR{Polchinski:1995mt}

\def\la{Affleck:1991yq}
\def\zyNew{Yin:prepare}
\def\zyBC{Yin:2002wz}
\def\ishibashi{Ishibashi:1989kg}
\def\clny{Callan:1987px}
\def\dms{DiFrancesco:1997nk}
\def\bcn{Bloete:1986qm}
\def\dhvw{Dixon:1985jw}

\def\thurston{Thurston:1977gttm}
\def\cftRev{\dms}
\def\appRev{ACGlesHouches:1989}

\begin{document}

\ifthenelse {\isundefined {\outputNonpreprintP}} {
% for noncompact only
	\begin{titlepage}

		\hfill
		\vbox{ 
		\halign{#\hfil     \cr   
		\thePreprintNumbers \cr
				   } 
			  }
		\ifthenelse {\isundefined {\draftFormatP}} {}
			{\hfill \verb$Id: qbs1.tex,v 1.10 2007/01/07 18:26:33 yin Exp $}
		\vspace*{20mm}
		\begin{center}
		
		{\Large {\bf  \thePreprintTitle}\\}
		\vspace*{15mm}
		{\theAuthorList\footnote{\theAuthorFootnote}}
		
		\vspace*{5mm}
		{\it {\theAddress}}\\
		
		\vspace{10mm}
		
		\end{center}
		
		\begin{abstract}
		\theAbstract
		\end{abstract}
		
		\vspace*{15mm}
		\flushleft \theDate
	
	\end{titlepage}
}
{\compactTitle}	% for compact only

%%%%%%

% start of content

%%%%%%

\sct {Introduction}
Conformal invariance in two dimensions\cite {\cftRev} is a powerful tool
for a wide range of physical problems where two-dimensional field
theories are relevant\cite {\appRev}, from condensed matter physics and
statistical mechanical models to the perturbative formulation of string
theory\cite {\polchinskiBook}. Although field theories are
most often formulated on a smooth manifold, in recent years it has
become evident that geometric singularities play an essential role both
in theory and application of conformal field theories.  By far the most 
studied type of singularity is boundary. Boundary conformal field theory
has been successfully applied to finite size effect\cite {\bcn}, Kondo
effect\cite {\la}, and D-branes and orientfolds via open
strings\cite {\polchinskiRR}.

An important object in boundary conformal field theory is boundary
state, which represents the geometric boundary algebraically as a
generalization of quantum state\cite {\clny, \ishibashi}.  It not only
encodes
physical information about a conformal boundary but also reveals deep
structure of the bulk theory\cite {\cardy, \bppz}.  A similar 
notion is cross cap state, which is relevant for unoriented
surfaces and open strings\cite {\as,\bh}.  

In addition to boundary there are other types of singularities 
that are very common in the real world: conical
defects and corners.  They appear naturally in the geometry of 
physical models\cite {\cp}, open string field theory\cite {\im}, 
and the possibility of rectangular open string\cite {\iim}.  
In this paper we present a general and
systematic approach to study such singularities using orbifolds. Conical
and corner defects can be as easily realized in orbifolds as boundaries
and cross caps.  In fact they are shown to be closely related and
studied together in an unified manner.
Generalizations of boundary states provide exact description
of singularities representative of those from the mathematical 
classification of orbifolds\cite {\thurston} and allows us to calculate
correlation functions.  We also analyze the degree of freedom localized
at the singularity and show that their spectra correspond to those of
the conformal fields on the bulk and boundary. We restrict the cases
considered here to specific orbifolds whose covering space is a sphere.
Technical detail and generality will appear separately.

%\sct {geometry and symmetry}

\sct {Quasi-boundary}
The simplest smooth two-dimensional surface of finite size is the
sphere.  Its isometry group is $O(3)$ and has three conjugacy classes
of involutions. Each such involution
generates one of the three
simplest types of nontrivial orbifolds whose covering space is the
sphere.  Let us parameterized the sphere with a complex coordinate $z$
using a stereographic projection.  These orbifolds are shown in \figRef
{Ztwo} with their ``primitive unit cell'' (i.e. fundamental domain)
shaded, borrowing a term from crystallography.  We have chosen it to be
the unit disc for all three cases.

\FigRef {Ztwo}.a represents the disc and \figRef {Ztwo}.b the cross cap.
The corresponding orbifold actions are respectively $z \to 1/\bar z$ and $z
\to - 1/\bar z$.  The former has a circle full of fixed points, $\abs {z} =
1$, which is the boundary of the disc; the latter has no fixed point, but 
diametrically opposite points on the unit circle are identified.

\ifthenelse {\isundefined {\outputNonpreprintP}} 
	{\newcommand {\picZtwo} {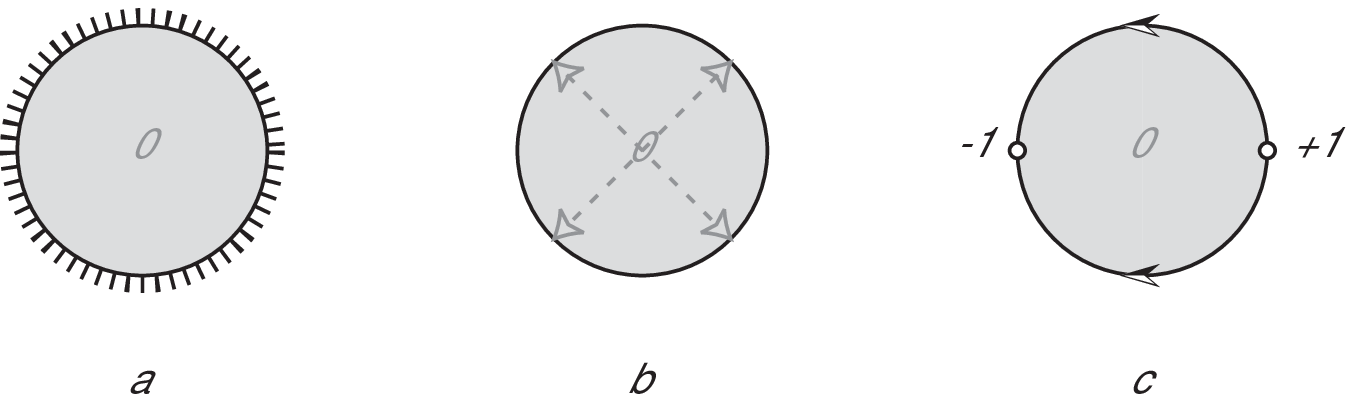}}
	{\newcommand {\picZtwo} {smallZ2.eps}}

\zyEpsf {\picZtwo}
		{Ztwo} 
		{$\ZZ_2$ orbifolds of sphere. (a): disc; (b): cross cap; 
		(c): bicorne cap with singularity at $z = \pm 1$.	}
%
%{	\zyEpsf {} {Ztwo} {		$\ZZ_2$ orbifolds of sphere. (a): disc; (b): cross cap; \\		(c): bicorne cap with singularity at $z = \pm 1$.	}}

\FigRef {Ztwo}.c represents the orbifold obtained with $z \to 1/z$. The
points on the unit circle are identified by a reflection across its
horizontal diameter, turning the circle into the line segment
$S^1/\ZZ_2$.  The two
ends of the line segment, $z = \pm 1$, are the fixed points of the
orbifold group. This orbifold has no boundary, but is singular at $z =
\pm 1$ because they have angular deficits of $\pi$.

We shall collectively call the boundary of a chosen primitive 
cell of an orbifold in its covering space its {\em quasi-boundary}.
In general, a quasi-boundary is not really a boundary of the orbifold. A
physical theory on a orbifold is first formulated on its covering
space and then the degree of freedom is restricted by choosing a
primitive cell on the covering space, which amounts to choosing the
quasi-boundary. Different choices of quasi-boundary in principle give
equivalent results, but in practice random choices would lead to
intractable calculation, while a judicious choice could allow efficient
or exact analysis.  This is why quasi-boundary is relevant and
important.  It plays a role similar to boundary, which is a special
case.

%\sct {other examples}
Combining the above three involutions appropriately leads to quotients 
of the sphere with 4-element
orbifold groups (\figRef {ZtwoZtwo}).  We
have again chosen to represent them all with the same geometric
quasi-boundary, now made up of the upper unit semi-circle and the interval
$\brak {-1, 1}$.  Their difference lies with the
local geometries of the quasi boundaries. \FigRef
{ZtwoZtwo}.a shows a $\ZZ_4$-orbifold generated by the map $z \to
\bar z$ and $z \to 1/\bar z$.  It has the topology of a disc so we call
it an open disc.  Here the quasi-boundary is a genuine
boundary of the open disc here, but it has two singular points $z = \pm
1$ because the boundaries intersect there at right angle instead of
$\pi$.  It should be pointed out that the physical import of the open
disc is not about an ``open'' CFT having boundary or an open string
getting absorbed by D-brane.% as sometimes suggested.  
Those would be
tautological and already accounted for by standard concepts in 
boundary CFT and open strings.  Rather, the novelty here is the
presence of singularities on the boundary of the worldsheet with
angular defects.

\ifthenelse {\isundefined {\outputNonpreprintP}} 
{
	\zyEpsf {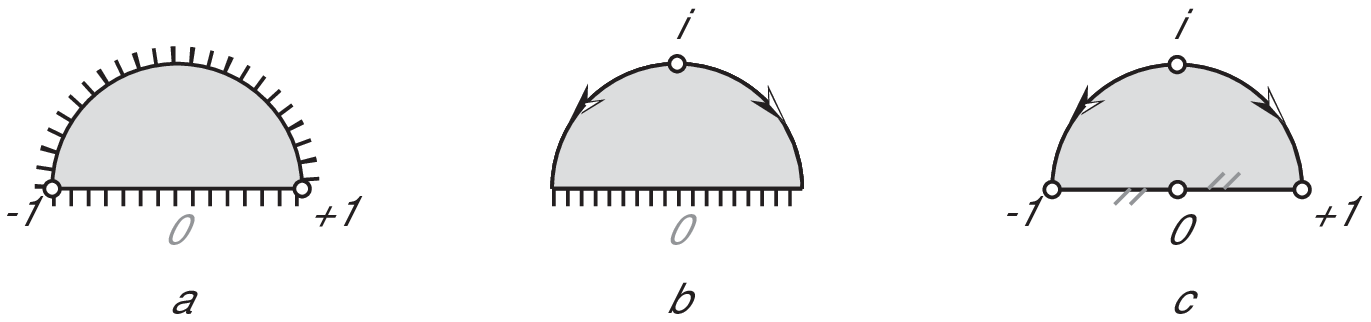} {ZtwoZtwo} {
		$\ZZ_2 \times \ZZ_2$ and $\ZZ_4$ orbifolds of sphere.  
		Singularities are marked with $\circ$.
		(a): open disc; (b): open bicorne cap; 
		(c): an orbifold with 3 fixed points.	
	}
}
{
	\zyEpsf {smallZ2xZ2.eps} {ZtwoZtwo} {
		$\ZZ_2 \times \ZZ_2$ and $\ZZ_4$ orbifolds of sphere.  
		Singularities are marked with $\circ$.
		(a): open disc; (b): open bicorne cap; 
		(c): an orbifold with 3 fixed points.	
	}
}

The $\ZZ_2 \times \ZZ_2$ orbifold group shown in \figRef {ZtwoZtwo}.b is
generated by $z \to \bar z$ and $z \to -1 / z$.  The line segment
$[-1, 1]$ on the real axis remains a bona fide boundary of the orbifold,
but the unit semi-circle on the upper complex plane does not.  Similar to
the bicorne cap, points on the semi-circle are subject to identification
by reflection with respect to the imaginary axis.  We call it the open
bicorne cap.  It also has the topology of a disc but with only one
singular point at $z = i$.

\FigRef {ZtwoZtwo}.c represents the $\ZZ_4$-orbifold of the sphere
generated by $z \to -z$ and $z \to 1/z$.  It has no boundary
at all and is topologically a sphere.  Both the semi-circle and the 
interval are subject to a $\ZZ_2$ identification as shown.  There are
three singular points, all with angular defect of $\pi$.  It has the
shape of a tortelli with three corners.

It is of course possible to consider more complicated orbifold groups as
well as surface of more complicated topology.  One can already describe
infinitely many other orbifolds by combining the varieties we have
already considered.  For example, any two quasi-boundaries from \figRef
{Ztwo} can be put together to describe a
$\ZZ_2$ orbifold of the torus.  On the other hand, a complete classification
of 2d orbifold has been made by categorizing the singular locus
\cite {\thurston}.  The latter fall into three families: the mirror
which is the fixed locus of reflection, elliptic points which are fixed
points of $Z_n$ rotation, and corner reflectors which are fixed point of
dihedral groups $D_n$.  Each family is already represented at least once
in \figRef {Ztwo} and \figRef {ZtwoZtwo}: the boundary is the mirror,
present in \figRef {Ztwo}.a, \figRef {ZtwoZtwo}.a and b; the singular
points in \figRef {Ztwo}.c, \figRef {ZtwoZtwo}.b and c are elliptic
points, which are angular defects in the bulk;  the singularities in
\figRef {ZtwoZtwo}.a are corner reflectors, which are angular defect on
boundary.  Hence the cases considered so far are representative.

\sct {Quasi-boundary state}
In all the cases discussed so far we have chosen so that $\abs {z} = 1$
overlaps with (part of) the quasi-boundary.  In radial quantization,
$\ln \abs {z}$ is the ``time.'' Therefore (part of) the quasi-boundary
sits at an instant in time.  It therefore corresponds to a state-like
object that we call {\em quasi-boundary state}. Note that there is no
requirement for a quasi-boundary state to be normalizable because
it does not in general represent a state of the physical system and
has no intrinsic probabilistic interpretation; it is a {\em state} in
the sense that the Hamiltonian can act on it. The description of such
systems is therefore turned
into a problem of finding the corresponding quasi-boundary state.  
The boundary and cross cap
states are known to be solutions\cite {\ishibashi} to sets of
equations involving the stress tensors and possibly other symmetry
currents.  Similarly, quasi-boundary state are solutions to a set of 
quasi-boundary conditions.

These conditions stipulate that the quasi-boundary state 
is preserved by the symmetries compatible with the orbifold condition.
Suppose that the corresponding symmetry charges are $Q_m$.  States have a
well defined
physical meaning only projectively, and the same should be true for
quasi-boundary states.  Therefore the general algebraic
statement for a quasi-boundary condition is
\beq	\label {eq:quasiBoundaryCondition}
	(Q_m - c_m) \BSket {} = 0.
\eeq
where $c_m$ is a set of c-numbers.

Let us first consider the conformal symmetry. The general procedure to
derive the symmetry that survives the orbifolding is similar to that
used in \cite {\zyBC}. The charges are $L_m - (\pm 1)^m \tL_{-m}$, $m
\in \ZZ$ for the disc (+) and cross cap (-) respectively. For the bicorne cap
the charges are $\bl_m \equiv L_m - L_{-m}$, and $\tbl_m \equiv \tL_m -
\tL_{-m}$, $m > 0$.  Note that unlike the previous two cases, there is
no mixing between the left and right chiral components, because its
orbifold group consists of pure rotation.  Although $\bl_m$ had not been
widely known, it originally appeared some time ago, probably in the
context of open string field theory\cite {\krr}.

For both the proper boundary state and the cross cap state, consistency
requires that $c_m = 0$.  However, for the bicorne cap, the
most general solutions are
\beqn	%\label {eq:generalBicorneBC}
	\bl_m - 2m (h_+ + (-1)^m h_-), \csp%\nono
	{\tbl}_m - 2m (\tlh_+ + (-1)^m \tlh_-).
\eeqn
Therefore the {\em bicorne cap state} is the solution of 
\beqar	\label {eq:bicorneCapBC}
	0 &=& \bracP { L_m - L_{-m} - 2m (h_+ + (-1)^m h_-) } \BSket {} \nono
	&=& \bracP { \tL_m - \tL_{-m} - 2m (\tlh_+ + (-1)^m \tlh_-) } \BSket {}
\eeqar
for $m > 0$.  The parameters $h_\pm$ and $\tlh_\pm$ have physical
meaning that will be discussed below along with excitations.

The above procedure for finding the symmetry charges can be applied to
all types of orbifold.  Here we briefly describe a couple more examples. For the
case of open disc (\figRef {ZtwoZtwo}.a), the $z \to \bar z$ action
reduces the conformal algebra to a single copy of Virasoro algebra $L_m$.
Then $z \to 1/\bar z$ further reduces it to $\bl_m$.  The boundary
condition is then just the first line of \eqr{bicorneCapBC}.  In a 
formal sense, this quasi-boundary state is the tensor square-root of
the bicorne cap state.  However, unlike the bicorne cap, in this case it
is in fact a bona fide boundary state, that is, of the open string
\cite {\im,\iim}, and two specific values of $h_\pm$ can be related to
Dirichlet and Neumann boundary conditions respectively.

For the case of open bicorne cap (\figRef {ZtwoZtwo}.c), the first step
is the same but then the single copy of Virasoro algebra is subject to $z
\to -1 / z$ instead.  This leads to the boundary condition
\beq
	\bracP {L_m - L_{-m} - 2m ( i^m h_+ + (-i)^m h_-)} \BSket {}
\eeq

%\sbsct {WZW model and higher symmetry}
One can also consider theories with bigger chiral algebra.  For such case,
analogous to boundary and cross cap states, one has the choice of
requiring the quasi-boundary (state) to preserve either just the relevant
conformal symmetry or also (part of) the other chiral symmetry.  In the
latter case, the conformal boundary conditions such as \eqr
{bicorneCapBC} is supplemented by additional conditions from the
generators of chiral symmetry.  Here we gave the result for the case of
affine Lie symmetries on the bicorne cap.  For detailed analysis and
other cases such as superconformal symmetry see \cite {\zyNew}.  Let
$J^a_m$ denote the affine current mode associated with some simple Lie
algebra $\lieg$.  The relevant boundary conditions are
\beq
	(J^i_m + U^i_j J^j_{-m}) \BSket {} = 0
\eeq
where $U$ is an involutary automorphism of $\lieg$, i.e. it satisfies
$U^2 = 1$ as well as $U^i_a U^j_b f^{ab}_k = f^{ij}_c U^c_k$ for
$\lieg$'s structure constant $f^{ij}_k$.  For bosonic string or
affine-U(1) currents $\alpha^\mu_m$, the conditions is
\beq
	(\alpha^\mu_m - R^\mu_\nu \alpha^\nu_{-m}) \BSket {} = 0
\eeq
where $R^\mu_\nu$ is an involution orthogonal with respect to the target
space metric.

\sct {Excitation at the singularity}
A physical system defined on a space with singularity introduces the
possibility of localized degree of freedom, which are excitations 
at the singularity.  A well known example is
boundary, which allows boundary fields describing changes of
boundary types.  Boundary fields must be distinguished from the
regular local scaling fields in the bulk, because the former are only
defined on a boundary component.  As a result they only transform under
the subalgebra of the conformal/chiral algebra preserved by the
boundary conditions.

The
bicorne cap state and the likes provide an exact description of
singularities. Therefore one expects to find a relation between these
quasi boundary states and excitations at the singularities. 
Here we derive the correct relation directly using \eqr {bicorneCapBC}.
Excitation at a singularity can be induced by a bulk field approaching
it. Let $\phi$
be a bulk primary field of conformal bi-weights $(\Delta, \tDelta)$ such
that
\beqn
	\psi_\pm \equiv \lim_{z->\pm1} \csp \lim_{\tz->\pm1} \csp
		(z \mp 1)^{\lambda_\pm} (\tz \mp 1)^{\tlambda_\pm} \csp \phi (z, \tz)
\eeqn
is finite. Then
\beqarn
	\comm {\bl_m, \psi_\pm} &=& 2m (\pm 1)^m d_\pm \psi_\pm, \\
%	\comm {\bl_m, \phi (-1)} = (-1)^m 2m \Delta \phi(1); \\
	\comm {\tbl_m, \psi_\pm} &=& 2m (\pm 1)^m \td_\pm \psi_\pm. \nonum
%	\comm {\tbl_m, \phi (-1)} = (-1)^m 2m \tilde \Delta \phi(1).
\eeqarn
Here $d_\pm = \Delta - \lambda_\pm$ and $\td_\pm = \tDelta -
\tlambda_\pm$ characterize the effect of $\psi_\pm$ on the
singularity.

Therefore $\psi_\pm$ act as shift operators for the parameters $h_\pm$
in \eqr {bicorneCapBC}. Since $\psi_\pm$ is localized at $z = \pm 1$, it
follows that $h_\pm$ characterize the singularity $z = \pm 1$
respectively. Each such point is therefore characterized by a pair of
parameters $(h, \tlh)$, and every excitation there is characterized by a
pair $(d, \td)$.  The excitation $\psi_\pm$ changes the singularity
by a shift $h \to h + d$, $\tlh \to \tlh + \td$.  This is
reminiscent of the relation between boundary and boundary operator\cite
{\cardy}.  The shift $d$ and $\td$ depend not only on the conform
weights of $\phi$ but also on the order of the zero/pole needed to
extract $\psi$.  It should be emphasized that $\lambda$ and $\tlambda$
depend on the specific theory and details of the singularity, so they
are not determined solely by $\Delta$ and $\tDelta$.  In
fact the complete information is encoded in a bulk-singularity expansion
that generalizes the bulk-bulk and bulk-boundary operator product
expansions.  Analogous result holds 
for the open bicorne.  For open boundary, the situation is similar 
except that instead of a chiral/anti-chiral pair, there
is just a single parameter associated with each corner reflector and its
excitation.

In boundary conform field theory, not only are the boundary operators
labeled by the conformal weights, but also the boundaries themselves.
This is because there is a type of boundary corresponding to the
identity field.  The same holds here.  Even though it might seem that
$h$ has no preferred value because it can be shifted by excitation, we
have found on general and reliable ground that it has a special value at
$- c/8$ that correspond to ``vacuum,'' c being the central charge.  For
unitary theories, this is a lower bound.
Therefore the spectra of excitations at elliptic and corner
singularities respectively correspond to those of the bulk and boundary
fields and have the meaning of scaling dimension.

\sct {Explicit solutions}
Quasi-boundary states not only provide an exact description of orbifold
singularity, but also allows explicit calculations in the operator
formalism.  Here we gave several examples of correlation functions of
vertex operators in the theory of $D$ free bosons (the Gaussian model):
\beqn
	\bracA {\, V_{k_1}(z_1,\tz_1) \ldots V_{k_n}(z_n,\tz_n) \,}
\eeqn
Detailed derivation will be given in \cite {\zyNew}.
For this theory solutions of quasi-boundary states can be explicitly constructed.  For example, 
\beq
	\bracP {\exp (\sum_m G_{\mu\nu} \alpha^\mu_{-m}\alpha^\nu_{-m}) \ket {0\,}}
	\otimes \ket {\chi\,}
\eeq
is a bicorne cap state.  Here $G_{\mu\nu}$ is the metric for the free boson,
$\ket {0\,}$ is the oscillator vacuum, and $\ket {\chi\,}$ is the
quantum state for the string center of mass. Using this state we find
the vertex operator correlation function on the bicorne cap to be
\beq	\label {eq:bicornCF}
	\prod_{1 \leq a < b \leq n} 
				\abs {\frac {z_a - z_b} {1 - z_a z_b}}^{2k_a \cdot k_b}
	\prod_{a=1}^n \abs {1-z_a^2}^{-k_a^2}
	2\pi \chi_k^*,
\eeq
where $k = \sum_{a=1}^n k_a$ and $\chi_k$ is the Fourier component of
the wave function $\chi(x)$ for momentum $k$.

For the open disc, the correlation function is
\beqar	\label {eq:openBCF}
	\prod_{1 \leq a < b \leq n} &&
				\abs {\frac {(z_a - z_b)(z_a - \tz_b)} 
							{(1 - z_a z_b)(1 - z_a \tz_b)}}^{2k_a \cdot k_b} \\
	\prod_{a=1}^n \,\,\,\,\,\, && 
		\abs {\frac {z_a - \tz_a}{(1-z_a^2)(1-\abs{z_a}^2)}}^{k_a^2}
	2\pi \chi_k^*. \nonum
\eeqar
And for the open bicorne cap, the correlation function is 
\beqar	\label {eq:openBicornCF}
	\prod_{1 \leq a < b \leq n} && 
				\abs {\frac {(z_a - z_b)(z_a - \tz_b)} 
							{(1 + z_a z_b)(1 + z_a \tz_b)}}^{2k_a \cdot k_b} \\
	\prod_{a=1}^n\,\,\,\,\,\,&& 
		\abs {\frac {z_a - \tz_a}{(1+z_a^2)(1+\abs{z_a}^2)}}^{k_a^2}
	2\pi \chi_k^*. \nonum
\eeqar

\sct {Conclusion}
We conclude this paper by remarking on the difference between the
spacetime orbifolds that appear in string theory, and general conformal
field theory on two-dimensional orbifolds considered here.  At the
formal level, while here we simply formulate the usual conformal field
theory on a given orbifold, spacetime orbifold comes out of the
procedure of ``orbifolding,'' where a discrete symmetry is gauged and at
the same time the so-called twisted sectors are added to obtain a
different theory\cite {\dhvw}.

More concretely, they differ in the way in which orbifold geometry,
especially its singularities, is represented in physical quantities.
For string theory, the worldsheet is smooth while the spacetime is
classically a singular orbifold.  However, the stringy
effect of the twisted sectors masks the singularity of the orbifold and
yield nonsingular amplitudes.  This is considered to be a positive
attribute of string theory because it modifies the small scale geometry
of spacetime.  By contrast, the zeros and poles in the correlation
functions (eq. \ref{eq:bicornCF}, \ref{eq:openBCF}, \ref{eq:openBicornCF})
clearly indicates the presence and location of singularity.
This does not mean
the theory is sick, but instead shows that singularities of the
underlying geometry are fully revealed by the field theory.

\ifthenelse {\isundefined {\draftFormatP}} 
{

}
{% dynamic reference using bibtex
	\ifthenelse {\isundefined {\outputNonpreprintP}} 
		{\bibliographystyle {unsrt}}
		{\bibliographystyle {apsrev}}
	
	\bibliography 
	{bib-qbs1,string-literature,misc_literature,polchinski,affleck,%
	cardy,nappi,dixon_L_J,diFrancesco}
}

\end{document}